\documentclass[review]{elsarticle}
\usepackage{a4wide}
\usepackage{natbib} 			
\usepackage{graphicx}
\usepackage{amsfonts}
\usepackage{amsmath}
\usepackage{amsthm}
\usepackage{booktabs}
\usepackage{epsfig}
\usepackage{mathrsfs}
\usepackage[colorlinks]{hyperref}
\hypersetup{citecolor=blue,linkcolor= blue}

\begin{document}
\title{
Measuring capital market efficiency: Global and local correlations structure}
\author{Ladislav Kristoufek}
\ead{kristoufek@ies-prague.org}
\address{Institute of Economic Studies, Charles University, Opletalova 26, 110 00, Prague, Czech Republic\\
Institute of Information Theory and Automation, Academy of Sciences of the Czech Republic, Pod Vodarenskou Vezi 4, Prague 8, 182 08}
\author{Miloslav Vosvrda}
\ead{vosvrda@utia.cas.cz}
\address{Institute of Economic Studies, Charles University, Opletalova 26, 110 00, Prague, Czech Republic\\
Institute of Information Theory and Automation, Academy of Sciences of the Czech Republic, Pod Vodarenskou Vezi 4, Prague 8, 182 08}

\begin{abstract}
We introduce a new measure for the capital market efficiency. The measure takes into consideration the correlation structure of the returns (long-term and short-term memory) and local herding behavior (fractal dimension). The efficiency measure is taken as a distance from an ideal efficient market situation. Methodology is applied to a portfolio of 41 stock indices. We find that the Japanese NIKKEI is the most efficient market. From geographical point of view, the more efficient markets are dominated by the European stock indices and the less efficient markets cover mainly Latin America, Asia and Oceania. The inefficiency is mainly driven by a local herding, i.e. a low fractal dimension.\\
\end{abstract}


\maketitle


\journal{Physica A}

\section{Introduction}

A concept of the capital market efficiency is a central notion in the financial markets theory \citep{Fama1970,Fama1991}. This notion is generally used for an ideal image of the capital market enabling to process relevant information to the fundamental price generation. If the relevant information to the fundamental price generation is completely processed by the capital market price mechanism, then such capital market is called to be efficient. Thus the capital market efficiency accentuates the informational efficiency of capital markets. A notion of the efficient capital market represents such capital market where prices on traded securities, e.g. stocks, bonds, or property, already reflect all available information and that investors are completely rational. Consequently, the notion of the efficient capital market represents fair game pattern. No investor can have an advantage in predicting a return on an asset price, since no one has access to information not already available to everyone. It means that investors in the efficient capital market cannot expect to achieve abnormal returns systematically. In other words, the capital market is efficient if the fluctuations of returns are unpredictable \citep{Fama1970,Fama1991,Malkiel2003}. 

Paradoxically, an achievement of the ideal efficient capital market, enabling efficient allocation of investments, brings about no activity of investors and no activity of speculators. Because real life experiences with capital markets have shown that there are investors who indeed have been beating the capital markets in long-term, discrepancies from the above mentioned ideal state are existent and thus worth analyzing.

Testing the efficiency of various capital markets in different regions is a popular topic in financial journals (e.g. \cite{Lam2011,Lim2008,Alexeev2011,Chong2011,Charles2011}). However, the hypothesis of market efficiency is standardly either rejected or not and markets are ranked quite infrequently. Moreover, the researchers majorly focus on a single method and comment on the results. And even further, the whole idea of testing or measuring capital market efficiency has been dealing with the joint-hypothesis problem (i.e. when we reject the efficiency of a specific market, it might be caused by a wrong assumption of the market's behavior) since its beginnings. This issue was also touched by Fama himself \citep{Fama1991}. In this paper, we try to bypass the problem by defining the efficient market as a martingale. We then analyze fractal dimension, and long-range and short-range dependence to describe and measure the efficiency of specific markets. 

Hurst exponent and a presence of long-term memory has been widely analyzed in recent years -- in stock indices \citep{DiMatteo2007,Kristoufek2010}, interest rates \citep{Cajueiro2007}, bonds \citep{Carbone2004}, exchange rates \citep{Vandewalle1997} and others. The results vary depending on asset type and on geographical situation as well. Statistically significant long-range dependence was detected in some individual NYSE-listed stocks \citep{Barkoulas1996}. Even though the series of developed markets usually posses only short or no memory, emerging markets exhibit a different behavior \citep{Wright2001,Podobnik2006}. Looking at a different frequency, a significantly long memory was found for weekly returns of a large number of Greek stocks \citep{Panas2001}. Cajueiro \& Tabak \citep{Cajueiro2004b} rank the markets according to their efficiency and suggest that Hong Kong stock exchange is the most efficient one followed by Chinese A type shares and Singapore and finally by Chinese B type shares, which indicates that liquidity and capital restrictions should be taken into consideration in efficiency testing and mainly interpretation.

We use the Hurst exponent $H$ and the fractal dimension $D$ to construct a new measure of market efficiency based on a deviation from the ideal state (the efficient market) from both local and global perspective. If the results based on different measures vary, we can further distinguish between local (herding) and global (correlations structure) effects. We use the fact that the measures are bounded and thus can be used to construct an informative norm representing the said deviation from the ideal state. The measure is estimated for 41 stock indices at different stages of development from the beginning of 2000 till the end of August 2011, i.e. the data set includes the DotCom bubble and its bursting as well as the current Global Financial Crisis.

The paper is structured as follows. In Section 2, we define the efficient capital market. Section 3 describes relationships between efficiency and the measures we use. In Section 4, we describe the methods used for the fractal dimension and Hurst exponent estimation. Section 5 covers the results and discusses the implications. Section 6 concludes. The main value of this paper lies in the fact that the proposed methodology bypasses the standard caveats of efficiency testing by building on the martingale definition of efficiency, using different methods and merging them into the efficiency measure. Such rather bold path leads to very interesting and also meaningful results.

\section{Capital market efficiency}
\label{EMH}

We use a triple $(\Omega, \mathscr{T}, P)$ for expressing a probability space and the expression $\mathbb{E}[X|\mathscr{T}]$ for the conditional expectations. Let $\{\omega \in \Omega\}$ be a set of elementary market situations. Let $\mathscr{T}$ be some $\sigma$-algebra of the subsets of $\Omega$, $P$ is a probability measure on $\mathscr{T}$ and $\Omega$ is an information set. This structure gives us all the machinery for static situations involving randomness. For dynamic situations, involving randomness over time, a sequence of  $\sigma$-algebras $\{\mathscr{T}_t, t \ge 0\}$ needs to be taken into consideration. Inclusion $\mathscr{T}_t \subset\mathscr{T}_{t+1}$ for all $t$ represents the information arriving in time $t$. Suppose all $\sigma$-algebras to be complete. Thus $\mathscr{T}_0$ represents initial information. On the other hand, a situation that all is known is represented by the expression $\mathscr{T}_{\infty}=\lim_{t \rightarrow \infty}\mathscr{T}_t$. Such a family $\{\mathscr{T}_{t\ge 0} \}$ is called a filtration; a probability space endowed with such a filtration, $(\Omega, \mathscr{T},\{\mathscr{T}_t\}, P)$, is also called a stochastic basis. 

Let $C=(\Omega, \mathscr{T},\{\mathscr{T}_t\}, P)$  be a capital market with distinguished flows $\{\mathscr{T}_{t\ge 0}\}$  of $\sigma$-algebras filtered probability space. We also call $\{\mathscr{T}_{t\ge 0}\}$ an information flow, and an expression $\{S_t\}_{t\ge 0}\in M$ is a security price process. The efficient market is then defined as follows:\\

\textit{A capital market $C=(\Omega, \mathscr{T},\{\mathscr{T}_t\}, P)$ is called efficient if there exists $P$ such that each security price sequence $S=\{S_t\}_{t\ge 0}$ is a $P$-martingale, i.e. the variables  are  $\mathscr{T}_t$-measurable and}

\begin{equation}
\mathbb{E}_P[|S_t|]<\infty, \mathbb{E}_P[S_{t+1}|\mathscr{T}_t]=S_t, t \ge 0
\end{equation} 


If a sequence $\{\xi_t\}_{t\ge 1}$ is the sequence of independent random variables such that $\mathbb{E}_P[|\xi_t|]<\infty$, $\mathbb{E}_P[\xi_t]=0$ for $t\ge1$, $\mathscr{T}_t^{\xi}=\sigma(\xi_1,\ldots,\xi_t)$, $\mathscr{T}_0^{\xi}=\{\emptyset, \Omega\}$, and $\mathscr{T}_t^{\xi} \subseteq \mathscr{T}_t$  then, evidently, the security price sequence  $S=\{S_t\}_{t\ge 0}$, where $S_t=\xi_1+\ldots+\xi_t$ for $t\ge 1$ and $S_0=0$, is a martingale with respect to $\mathscr{T}^{\xi}=\{\mathscr{T}_t^{\xi}\}_{t\ge0}$, and 

\begin{equation}
\mathbb{E}_P\left[S_{t+1}|\mathscr{T}_t\right]=S_t+\mathbb{E}_P\left[\xi_{t+1}|\mathscr{T}_t\right].
\end{equation}

If a sequence  $S=\{S_t\}_{t\ge 0}$ is a martingale with respect to the filtration $\{\mathscr{T}_{t\ge0}\}$ and $S_t=\xi_1+\ldots+\xi_t$ for $t\ge 1$ with $S_0=0$, then $\{\xi_t\}_{t\ge1}$   is a martingale difference, i.e. $\xi_t$ is $\mathscr{T}_t$-measurable, $\mathbb{E}_P\left[|\xi_t|\right]<\infty$ and $\mathbb{E}_P\left[\xi_t|\mathscr{T}_{t-1}\right]=0$.

Thus in words, the capital market efficiency is represented by the martingale property of the security price processes. Note that this feature is primarily connected to uncorrelated returns of the price series\footnote{Note that security price process $S=\{S_t\}_{t\ge 0}$ can be taken either as a simple price or logarithmic price process. In our application, we use the more standard approach, i.e. the logarithmic prices.}. Compared to the random-walk-based efficiency, the martingale is more general and does not assume the series to be locally stationary (homoskedastic), which would be quite unrealistic for the financial time series. Nevertheless, the martingale assumption gives enough information about the expected Hurst exponent and fractal dimension. 
Note that our information set $\Omega$ contains only the prices of the analyzed indices so that we test and measure the weak form of the capital market efficiency.

\section{Relationship between efficient market, fractal dimension and long memory}

Traditionally, time series of martingale fluctuations are described as being generated by the Gaussian noise (or the Brownian motion for an integrated process) and space of many time series of such fluctuations to be generated by some mixing mechanism of the Gaussian noise interspersed with L\'{e}vy jumps (or again as the Brownian motion interspersed with L\'{e}vy flights when talking about the integrated processes). Describing a theoretical model of the efficient capital market with Brownian motion, we asymptotically obtain a normal (Gaussian) distribution of returns and an empirical  distribution of returns will be asymptotically very close to the normal distribution. In a multidimensional theoretical model of the efficient capital market with the multidimensional Brownian motion, we asymptotically obtain a multidimensional normal distribution of returns and multidimensional empirical distribution of returns will be asymptotically very close to the multidimensional normal distribution. However, the martingale definition of the efficient capital market as outlined in Section \ref{EMH} does not necessarily require Brownian motion (practically any integrated process of serially uncorrelated and finite variance increments suffices). Nevertheless, we shall see that in many cases, the efficient market leads to the Brownian motion.  

A measure of roughness $D$ of the $n$-dimensional sphere is called a fractal dimension. The fractal dimension $D\in \langle n,n+1)$ for hyperplane $\mathbb{R}^{n+1}$ is a local characteristic of the time series. Long-term memory in the time series of fluctuations on the capital market is connected to the power law of autocorrelations. This effect is usually called the Hurst effect and is measured by the Hurst exponent $H$ \citep{Hurst1951}. Long-term memory is a global characteristic of the series. Without imposing further assumptions about the underlying process, $D$ and $H$ are independent. For self-affine processes, it holds that $D+H=n+1$. For a univariate case that means $D+H=2$ \citep{Gneiting2004}. In context of the capital markets, such relation implies that local behavior (such as a herding behavior or fear) is reflected in the global characteristics (such as significant autocorrelations or high volatility).

If we assume that the local behavior is at least partially projected into the global features of the market, then persistence is connected to the low fractal dimension $D\in(1,1.5)$ and the higher fractal dimension $D\in (1.5, 2)$ is connected to anti-persistent processes. If the process is characterized by Hurst exponent close to 0.5 and the fractal dimension close to 1.5, it should have no correlation structure. 

\section{Methodology}


\subsection{Long-range dependence, Hurst exponent and market efficiency}

Long-range dependence (or long-term memory) is a feature of time series' autocorrelations. If the series is long-range dependent, the autocorrelation function $\rho(k)=\mathbb{E}[(X_t-\mathbb{E}[X_t])(X_{t-k}-\mathbb{E}[X_{t}])]/\mathbb{E}[(X_t-\mathbb{E}[X_t])^2]$, where $X_t$ is a stationary process, decays asymptotically hyperbolically, i.e. $\rho(k) \propto k^{2H-2}$ for $k \rightarrow \infty$. Therefore, the behavior of the series has infinite memory, i.e. the shocks in a very distant past may have a significant effect on the today's behavior. Such behavior is in violation with the definition of an efficient market (see Section \ref{EMH}) because it allows for arbitrage as shown by \cite{Mandelbrot1968}. The relationship between long-term memory, predictability and potential efficiency has been discussed in numerous studies \citep{Cajueiro2004,Cajueiro2004a,Cajueiro2005,DiMatteo2003,DiMatteo2005,Eom2008} and in many cases, the less developed markets were characterized by signs of long-term memory. Moreover, the time-dependent Hurst exponent has been used several times to describe various phases of the financial markets and its connection to the efficient markets \citep{Morales2012,Kristoufek2012a}.

Characteristic measure of long-range dependence is Hurst exponent $H$, which ranges between 0 and 1. For $H=0.5$, the series is considered serially uncorrelated or short-term correlated, whereas for $H \ne 0.5$, we consider it long-range correlated. More specifically, $H>0.5$ indicates persistence or positive long-term memory, which is usually interpreted in a way that a positive increment of the series is more likely to be followed by another increment and vice versa. Inversely, $H<0.5$ indicates anti-persistence or negative long-term memory which is connected with more frequent switching of the increments and decrements than would be observed for a random process. Both types can be exploited to obtain abnormal returns on the market since the fluctuations are predictable \citep{Mandelbrot1968}. However, it needs to be noted that for financial time series, even an uncorrelated series can yield $H\ne 0.5$ due to several reasons such as heteroskedasticity, short-term memory and fat tails. The effects of these are discussed in various papers, e.g. \citep{Kristoufek2012,Barunik2010,Barunik2012}.

In our analysis, we apply various methods to estimate the Hurst exponent $H$ -- detrended fluctuation analysis \citep{Peng1993,Peng1994,Kantelhardt2002}, detrending moving average \citep{Alessio2002}, 
and height-height correlation analysis (also known as the generalized Hurst exponent approach) \citep{DiMatteo2003,Alvarez-Ramirez2002,Barabasi1991}.



\subsubsection{Detrended fluctuation analysis}

Detrended fluctuation analysis (DFA) \citep{Peng1993,Peng1994} is based on the scaling of variances of the detrended series. We split the series into boxes of length $s$ and estimate a polynomial fit of the profile $\overline{X_{t,s}}$. The detrended series is constructed as $\overline{Y_t}=X_t-\overline{X_{t,s}}$. Fluctuations $F_{DFA}^2(s)$, which are defined as an average of the mean squared error of the polynomial fit over all boxes with length $s$, scale according to $F_{DFA}^2(s)\propto s^{2H}$ \citep{Kantelhardt2002}. Note that we use $s_{min}=5$ and $s_{max}=T/5$.  

\subsubsection{Detrending moving average}

Detrending moving average (DMA) \citep{Alessio2002} is based on a moving average filtering. For a window length $\lambda$, we construct a centered moving average $\overline{X_t}$ for each possible point of $X_t$. Similarly to DFA, we define the fluctuation $F_{DMA}^2(\lambda)$ as a mean squared error between $X_t$ and $\overline{X_{t,\lambda}}$, which scales as $F_{DMA}^2(\lambda)\propto \lambda^{2H}$. As we are using the centered moving average, we use $\lambda_{min}=3$ and $\lambda_{max}=21$ with a step of 2. 

\subsubsection{Height-height correlation analysis}

Height-height correlation analysis (HHCA) \citep{Barabasi1991}, also known as the generalized Hurst exponent approach (GHE) \citep{DiMatteo2003}, is based on scaling of height correlation function of series $X_t$ with time resolution $\nu$ and $t=\nu,2\nu,\ldots,\nu\lfloor T/\nu\rfloor$ (where $\lfloor \rfloor$ is a lower integer operator). The height correlation function of the second order for series $X_t$ is defined as $K_2(\tau)=\sum_{\tau=1}^{\lfloor T/\nu \rfloor}{|X_{t+\tau}-X_t|^2/\lfloor T/\nu \rfloor}$ where $\tau$ ranges between $\nu=\tau_{min},\ldots,\tau_{max}$. In our application, we use $\tau_{min}=1$ and $\tau_{max}$ varies between 5 and 20. This way, we obtain more estimates of the Hurst exponent and take their average as our best estimate, i.e. we apply jackknife which is standardly done for this method \citep{DiMatteo2005}. 

\subsection{Efficient market and fractal dimension}

As already mentioned in the previous section, fractal dimension $D$ is a measure of roughness on $\mathbb{R}^{n+1}$ for $n$-dimensional time series (the additional dimension in $n+1$ represents time). For a random series, the fractal dimension $D=1.5$. In the real markets, short-term trends usually occur and are connected to so-called "bear" (declining market with a negative mood) and "bull" (increasing market with a positive mood) markets. These short-term episodes on the market are not reflected into its global characteristics since these would quickly vanish due to arbitrage potential. Nevertheless, the episodes cause roughening of the series which is in turn reflected in the deviation of $D$ from 1.5. If the market is characterized by "local persistence", i.e. short-term trends, the fractal dimension should be below 1.5 as the surface of the series becomes smoother. Reversely, if the market is dominated by "local anti-persistence", i.e. short-term bursts of volatility, the fractal dimension should be higher than 1.5 as the surface of the series is more coarsened. Of course, these effects can both be present on the market.

Fractal dimension is connected to the fractal nature of a geometric object and is connected to the Hausdorff dimension \citep{Mandelbrot1982} and is usually estimated through such dimension for a graphical object or texture \citep{Neves2011}. However, we cannot estimate the fractal dimension $D$ this way for a univariate time series and it is needed to use an alternative estimators used for time series -- periodogram estimator \citep{Chan1996}, Genton estimator \citep{Genton1998}, Hall-Wood estimator \citep{Hall1993} and wavelet-based estimator \citep{Gneiting2010,Percival2000}. Detailed description of the methods is given in Gneiting \textit{et al.} \citep{Gneiting2010} and references therein.

\subsection{Capital market efficiency measure}

For a construction of capital market efficiency measure $EI$, we use the fact that both fractal dimension $D$ and Hurst exponent $H$ are bounded. Hurst exponent for stationary series is defined on interval $\langle0,1)$ and fractal dimension for a univariate case is defined on $\langle1,2)$. For each measure, the value for an efficient market lies in the center of its support, i.e. $H=0.5$ and $D=1.5$. Various estimates of fractal dimension $\hat{D}$ are obtained as defined earlier in the text. For the estimated Hurst exponent $\hat{H}$, we use estimates based DFA, DMA and HHCA but for DFA and HHCA, we use also the alternatives. DFA estimate is calculated for both linear and quadratic trend filtering. HHCA/GHE is then based on definition of both Barabasi \textit{et al.} \cite{Barabasi1991} and Di Matteo \cite{DiMatteo2007}. The DFA estimate is taken as an average of the two alternatives and the same for HHCA. Such a procedure is chosen because each of the methods is better suited for different types of processes both for Hurst exponent and fractal dimension estimation \citep{Weron2002,Barunik2010,Kristoufek2010a,Gneiting2010}. It is needed to note that Hurst exponent estimators are usually biased by a presence of short-term memory in the underlying process \citep{Kristoufek2012,Barunik2012}. However, this is not an issue for the proposed efficiency measure because when short-memory (which is of course a form of inefficiency as well) biases Hurst exponent estimate, it is in turn reflected in the efficiency measure. To further control for short-range dependence in the analyzed series, we also include the first order sample autocorrelation $\widehat{\rho(1)}$ into $EI$. Note that $\rho(1)$ ranges between -1 (perfectly anti-correlated) and 1 (perfectly correlated) and thus has a range of 2 which needs to be controlled for. Therefore, $EI$ is based on 8 estimates (4 estimates of fractal dimension, 3 estimates of Hurst exponent and 1 estimate of the first order autocorrelation). For our specific case, $EI$ is defined as
\begin{equation}
\begin{split}
EI=[\sum_{i=1}^{8}(\widehat{H_{DFA}}-0.5)^2+(\widehat{H_{DMA}}-0.5)^2+(\widehat{H_{HHCA}}-0.5)^2+\\
(\widehat{D_{P}}-1.5)^2+(\widehat{D_{W}}-1.5)^2+(\widehat{D_{G}}-1.5)^2+(\widehat{D_{HW}}-1.5)^2+(\widehat{\rho(1)}/2)^2]^{\frac{1}{2}},
\end{split}
\end{equation}

where $\widehat{H_{DFA}}$, $\widehat{H_{DMA}}$ and $\widehat{H_{HHCA}}$ are estimated Hurst exponents based on DFA, DMA and HHCA/GHE, respectively, $\widehat{D_{P}}$, $\widehat{D_{W}}$, $\widehat{D_{G}}$ and $\widehat{D_{HW}}$ are estimated fractal dimensions based on periodogram method, wavelets, Genton method and Hall-Wood method, respectively, and $\widehat{\rho(1)}$ is the sample first order autocorrelation.

We can generalize $EI$ for $n$ measures so that

\begin{equation}
EI=\sqrt{{\sum_{i=1}^n{\left(\frac{\widehat{M_i}-M_i^{\ast}}{R_i}\right)^2}}},
\end{equation}

where $M_i$ is the $i$th measure of efficiency (Hurst exponents $H$, fractal dimensions $D$ and the first order autocorrelation $\rho(1)$ in our case), $\widehat{M_i}$ is an estimate of the $i$th measure, $M_i^{\ast}$ is an expected value of the $i$th measure for the efficient market and $R_i$ is a range of the $i$th measure. As ranges of different measures may vary, we standardize them so that the range is equal to one, implying a unit cube as a resulting space. For the efficient market, we have $EI=0$, and for the least efficient market, we have $EI=\frac{\sqrt{n}}{2}$, where $n$ is a number of measures taken into consideration. Therefore, the efficiency index is defined on a unit $n$-dimensional cube with the efficient market in the center, i.e $EI=0$ for the efficient market. 

\begin{figure}[htbp]
\center
\begin{tabular}{cc}
\includegraphics[width=6in]{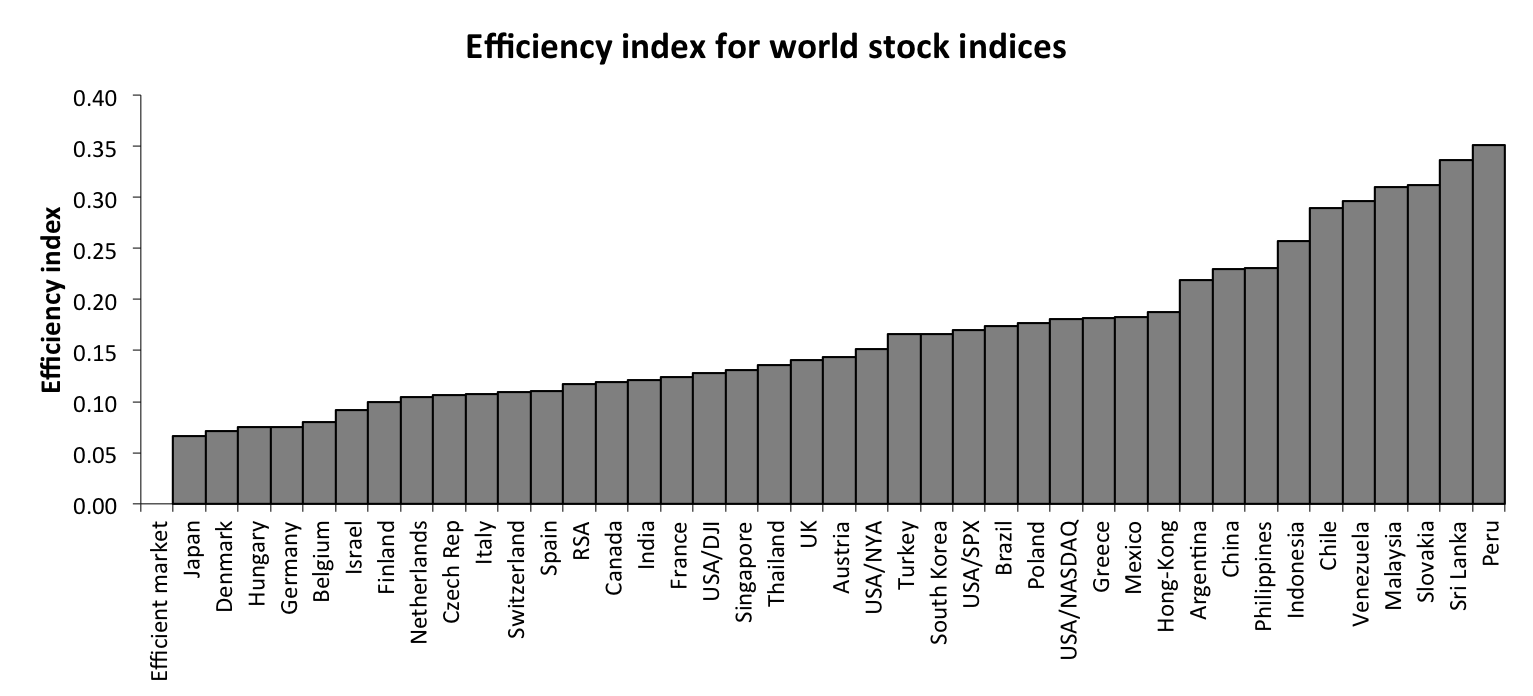}\\
\end{tabular}
\caption{\footnotesize\textit{Efficiency index.}\label{EI}}
\end{figure}

\section{Results and discussion}

We analyze efficiency of 41 stock market indices, which are described in Table \ref{tab1}. The dataset covers indices from the North and Latin America, Western Europe, Eastern Europe, Asia, Oceania and Africa and it has been obtained from dukascopy.com public database. The analyzed period ranges from the beginning of 2000 to the end of August 2011 (except of the indices which were founded later than 2000). The period thus includes each of the years of relatively stable growth, years of stable decrease after the DotCom bubble burst as well as the current crisis (crises). In Table \ref{tab2}, basic descriptive statistics of the logarithmic close/close returns are mentioned. Apart from the basic statistics (average, minimum, maximum, standard deviation, skewness and excess kurtosis), we show the KPSS statistics for the series. According to this test, all the series but CSE (Sri Lanka index) are weakly stationary. We do not show the results for ADF and PP unit-root tests since the test strongly rejects unit-root for all the indices with p-values practically equal to zero. 

Let us now turn to the results. The results for efficiency index $EI$ are summarized in Fig. \ref{EI}. In the figure, the indices are sorted so that the more to the left the index is, the more efficient it is. The most efficient market turns out to be the Japanese NIKKEI and the least efficient is the Peruvian IGRA. The "Top 5" of the most efficient indices contains the Japanese NIKKEI, Danish KFX, Hungarian BUX, German DAX and Belgian BEL20. From the opposite side, the "Bottom 5" includes Venezuelan IBC, Malaysian KLSE, Slovakian SAX, CSE of Sri Lanka and the already mentioned Peruvian IGRA. These are the results which might be labeled as relatively expected. However, when we look at the other indices, we can find more interesting results. First, the efficiency of the indices is geographically-dependent. For the more efficient half of the analyzed markets, there 13 European, 5 Asian, 2 North American and 1 African stock exchanges. The less efficient half contains 6 Latin American, 4 Asian, 4 European, 3 North American and 3 Oceanian. Therefore, the more efficient markets are dominated by European stock exchanges while the less efficient markets are dominated by Latin America, Asia and Oceania. Probably the most interesting are the performances of the US stock exchanges, which all score around the middle of the ranking. However, this should not be very surprising as the analyzed period contains both the DotCom bubble and the Global Financial Crisis of the late 2000s, which most severely hit the US markets and can be evidently taken as a source of inefficiency. Out of the four analyzed US indices, the most efficient turns out to be DJI and the least efficient is NASDAQ. Note that the most efficient US index, i.e. DJI, is less efficient than the biggest EU index -- German DAX. British FTSE, which can be considered as tightly connected to the US indices ranks very similarly to the US indices, in the middle of the ranking. To be able to comment on the sources of inefficiency, we now focus on the separate measures.

\begin{figure}[htbp]
\center
\begin{tabular}{cc}
\includegraphics[width=6in]{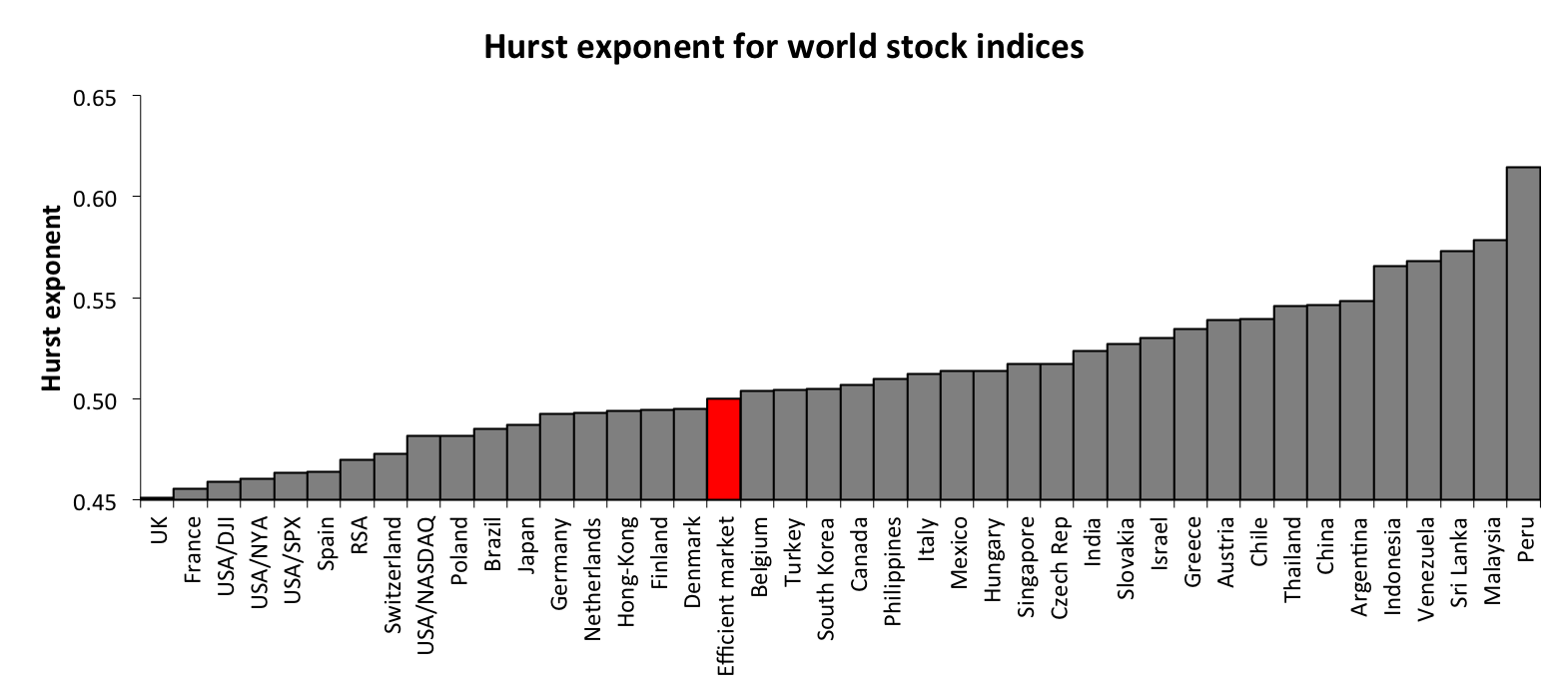}\\
\end{tabular}
\caption{\footnotesize\textit{Hurst exponent.}\label{H}}
\end{figure}

Focusing on the Hurst exponent estimates, we observe that the deviations from the ideal $H=0.5$ are not that severe for majority of the indices and generally range between 0.45 and 0.55. In Fig. \ref{H}, we show the average values of the Hurst exponent estimates. Interestingly, we find that majority of the most developed markets (but not necessarily the most efficient as shown above) are below the $H=0.5$ level, which is in hand with the results of Di Matteo \textit{et al.} \citep{DiMatteo2005}. Generally, we don't observe strong global correlation structure in the processes. This is, however, not a big surprise as the existence of strong global auto-correlations would lead to arbitrage opportunities, which are not likely to last for long taking into account the computerized trading in the current markets. The indices with the highest Hurst exponents are the same as the least efficient markets. From these, the Peruvian IGRA shows higher persistence than the non-stationary CSE of Sri Lanka. Results for the first order autocorrelation are very similar for all indices and close to zero so that we do not present them.    
   
\begin{figure}[htbp]
\center
\begin{tabular}{cc}
\includegraphics[width=6in]{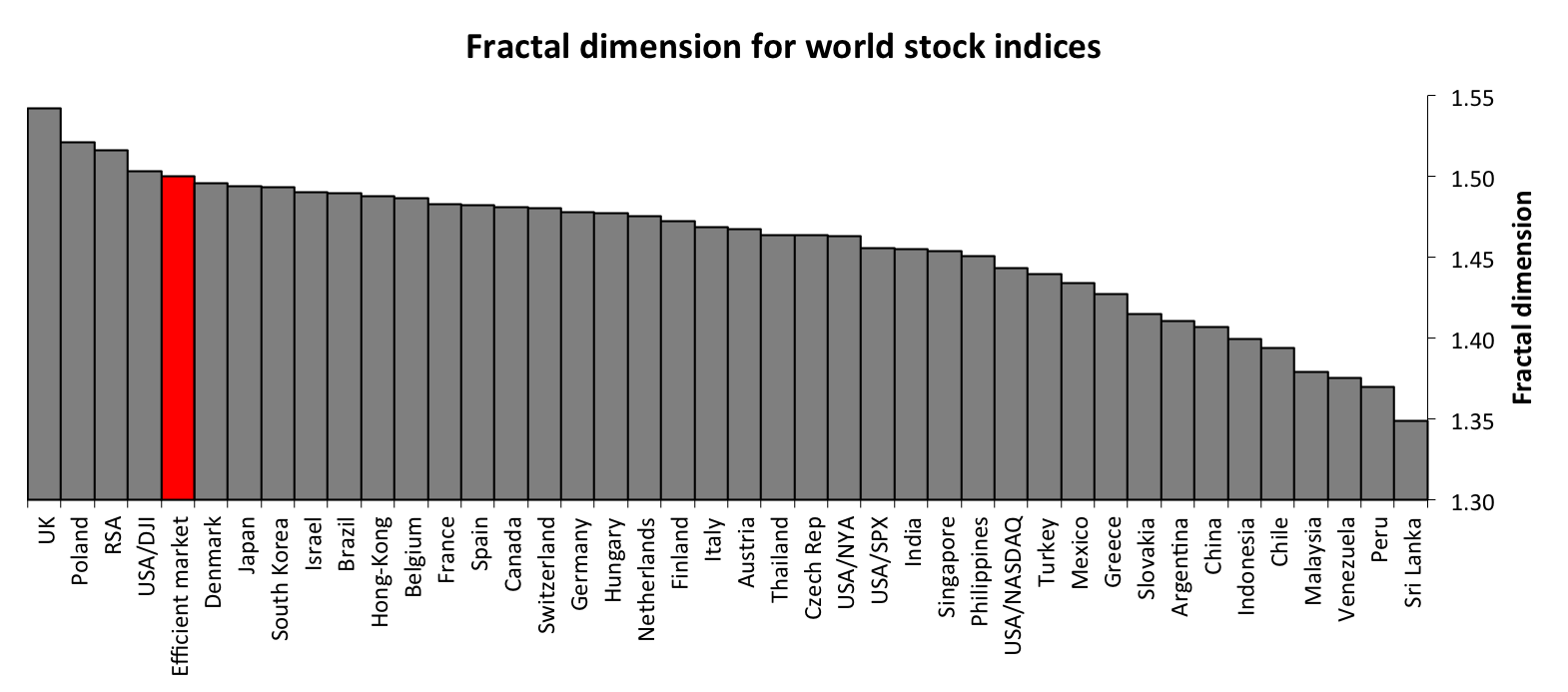}\\
\end{tabular}
\caption{\footnotesize\textit{Fractal dimension.}\label{D}}
\end{figure}

\begin{figure}[htbp]
\center
\begin{tabular}{cc}
\includegraphics[width=6in]{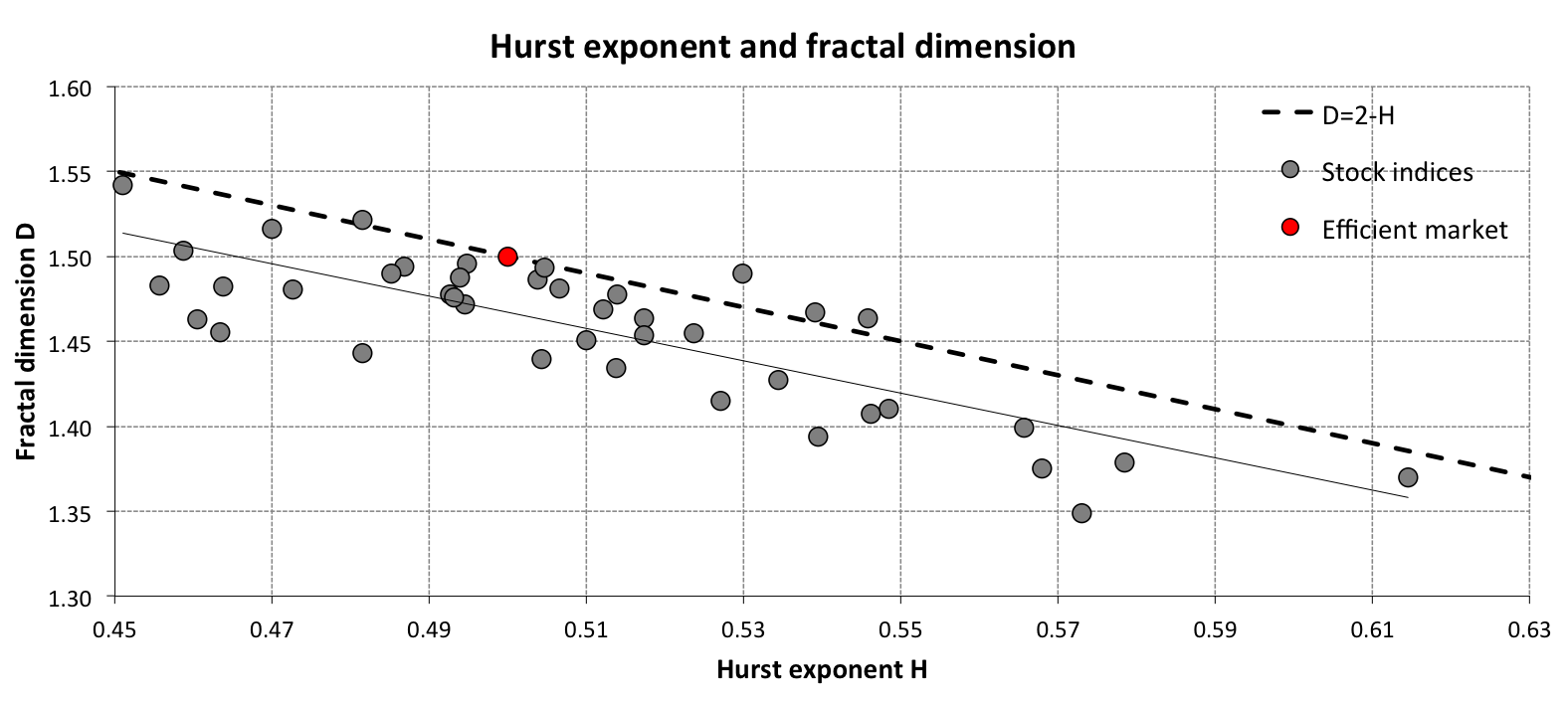}\\
\end{tabular}
\caption{\footnotesize\textit{Relationship between Hurst exponent and fractal dimension.}\label{DH_rel}}
\end{figure}

The results for fractal dimension $D$ (Fig.~\ref{D}) tell a more informative story. We observe that majority of the indices are characterized by the fractal dimension below 1.5, which indicates local persistence. For several indices, we observe the opposite. These are British FTSE, Polish WIG20, JSE of the Republic South Africa and DJI of the USA. Again the least efficient markets show the biggest deviation from $D=1.5$ and fall below 1.4. These markets thus experience relatively strong short-term trends and are at least partially predictable in the short term. Such a correlation structure even translates into global perspective as the markets with the lowest fractal dimension are also the markets with highest Hurst exponent. The general relationship between $H$ and $D$ for the analyzed indices is illustrated in Fig.~\ref{DH_rel}. We observe that even though $D$ and $H$ are not independent, the relationship is not exactly of form $D=2-H$ of self-affine processes, yet it is very close to it. Specifically, the estimated relationship is $\widehat{D}=1.94-0.95\widehat{H}$ with $R^2=0.66$ which implies that fractal dimensions $D$ are paired with lower Hurst exponents $H$ than for the self-affine processes. This is exactly in hand with an economic interpretation that the local herding and crowd behavior are short-lasting and are thus not reflected in the global measure of the Hurst exponent $H$ perfectly but only partially. Such a result obviously makes sense in the stock markets where such a strong global dependence would lead to profitable opportunities which would vanish due to the interaction between supply and demand in the market. Nonetheless, the deviations of $D$ and $H$ from their efficient market values for the least efficient markets show that not only the profit opportunities are important for investors but also an institutional framework such as legal and regulation issues as well as liquidity issues. Otherwise, the profit opportunities even for the least efficient markets would vanish quickly.

\begin{figure}[htbp]
\center
\begin{tabular}{cc}
\includegraphics[width=6in]{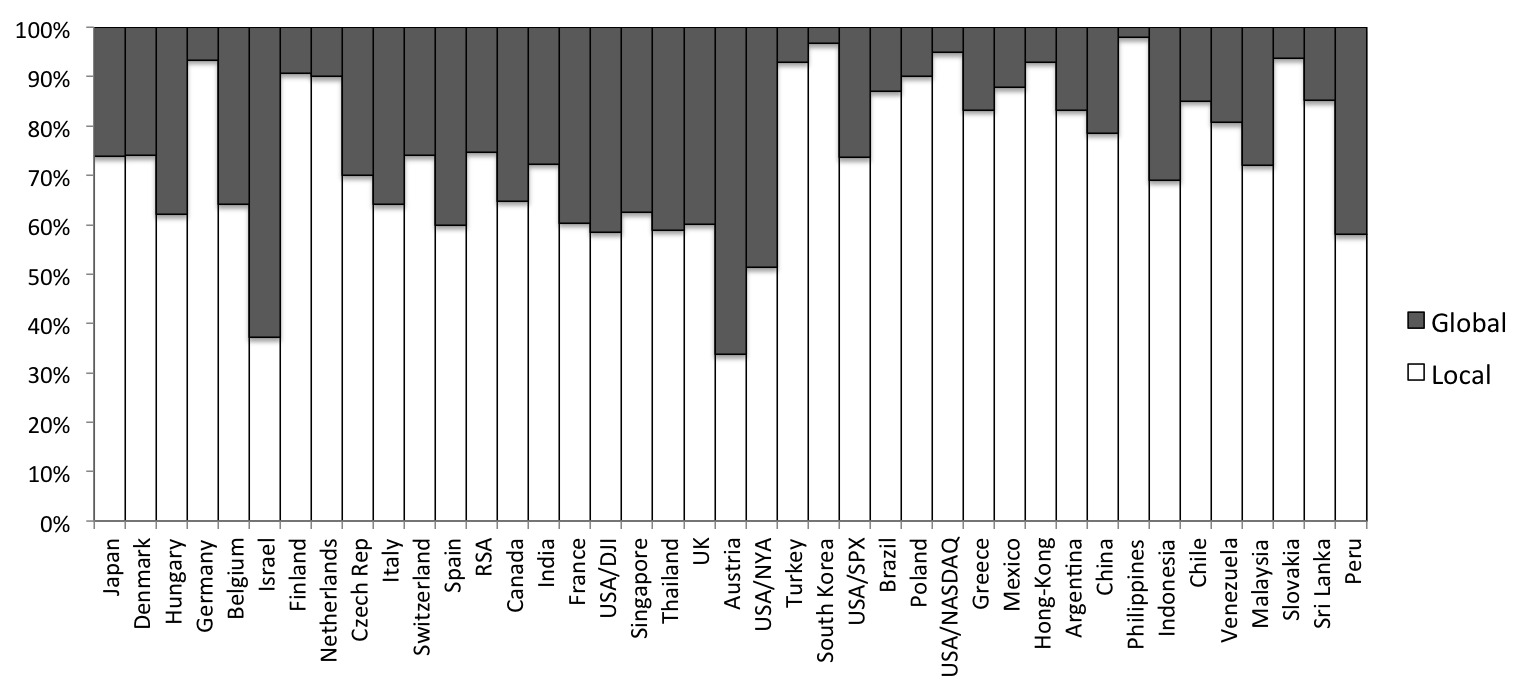}\\
\end{tabular}
\caption{\footnotesize\textit{Separation between local and global inefficiencies from the most efficient market on the left to the least efficient market on the right.}\label{LocalGlobal}}
\end{figure}

To further illustrate different the effects of local and global inefficiencies into the total $EI$ measure, we present Fig. \ref{LocalGlobal}. We can see that for majority of indices, the local inefficiencies (deviation of the fractal dimension from 1.5) dominate the global inefficiencies (short-term and long-term memory). Interestingly for the least efficient markets, the dominance is the most evident. There are two exceptions for which the global inefficiencies dominate and these are Israeli TA100 and Austrian ATX. For the more efficient markets, the proportion of local inefficiencies is around 40\% and for the less efficient markets, it is around 80\% of the total.

\section{Conclusions}

We have introduced a novel approach to measuring capital market efficiency. With a use of bounded measures of dynamic systems connected to a standard martingale definition of the capital market efficiency, we constructed a vector containing long-term memory, short-term memory and fractal dimension measures. The efficiency index $EI$ is calculated as a simple norm of this vector from its ideal efficient case. Therefore, a distance of a specific market situation from a centre of an $n$-dimensional cube is taken as a measure of efficiency. The further the market is from the ideal state, the less efficient it is. Such procedure can be easily generalized to more bounded efficiency quantities.

Applying the methodology on a set of 41 stock indices in period between 2000 and 2011, we found that the Japanese NIKKEI is the most efficient market. From geographical point of view, the more efficient markets are dominated by European stock indices and the less efficient markets cover mainly the Latin America, Asia and Oceania. More specifically, the least efficient markets are Venezuelan IBC, Malaysian KLSE, Slovakian SAX, CSE of Sri Lanka and Peruvian IGRA (the most inefficient stock market in the analyzed set). We also found that the local characteristics of the series (crowd and herding behavior) partially translate into the global characteristics (correlation structure). Moreover, the local inefficiencies in general dominate the total inefficiency for strong majority of the indices.

\section*{Acknowledgements}
This paper was written with the support of the Czech Science Foundation project No. P402/12/G097 ãDYME Ð Dynamic Models in EconomicsÒ. We also thank four anonymous referees for valuable comments which helped to improve the paper significantly.


\begin{table}[htbp]
\centering
\caption{List of the analyzed indices}
\label{tab1}
\footnotesize
\begin{tabular}{||c|c|c||}
\toprule \toprule
Ticker&Index&Country\\
\midrule \midrule
AEX&Amsterdam Exchange Index&Netherlands\\
ASE&Athens Stock Exchange General Index&Greece\\
ATX&Austrian Traded Index&Austria\\
BEL20&Euronext Brussels Index&Belgium\\
BSE&Bombay Stock Exchange Index&India\\
BUSP&Bovespa Brasil Sao Paulo Stock Exchange Index&Brasil\\
BUX&Budapest Stock Exchange Index&Hungary\\
CAC&Euronext Paris Bourse Index&France\\
CSE&Chittagong Stock Exchange Index&Sri Lanka\\
DAX&Deutscher Aktien Index&Germany\\
DJI&Dow Jones Industrial Average Index&USA\\
FTSE&Financial Times Stock Exchange 100 Index&UK\\
HEX&OMX Helsinki Index&Finland\\
HSI&Hang Seng Index&Hong-Kong\\
IBC&Caracas Stock Exchange Index&Venezuela\\
IGBM&Madrid Stock Exchange General Index&Spain\\
IGRA&Peru Stock Market Index&Peru\\
IPC&Indice de Precios y Cotizaciones&Mexico\\
IPSA&Santiago Stock Exchange Index&Chile\\
JKSE&Jakarta Composite Index&Indonesia\\
JSE&Africa All Share Index&RSA\\
KFX&Copenhagen Stock Exchange Index&Denmark\\
KLSE&Bursa Malaysia Index&Malaysia\\
KS11&KOSPI Composite Index&South Korea\\
MERVAL&Mercado de Valores Index&Argentina\\
MIBTEL&Borsa Italiana Index&Italy\\
NASD&NASDAQ Composite Index&USA\\
NIKKEI&NIKKEI 225 Index&Japan\\
NYA&NYSE Composite Index&USA\\
PSE&Philippine Stock Exchange Index&Philippines\\ 
PX&Prague Stock Exchange Index&Czech Republic\\
SAX&Slovakia Stock Exchange Index&Slovakia\\
SET&Stock Exchange of Thailand Index&Thailand\\
SPX&Standard \& Poor's 500 Index&USA\\
SSEC&Shanghai Composite Index&China\\
SSMI&Swiss Market Index&Switzerland\\
STRAITS&Straits Times Index&Singapore\\
TA100&Tel Aviv 100 Index&Israel\\
TSE&Toronto Stock Exchange TSE 300 Index&Canada\\
WIG20&Warsaw Stock Exchange WIG 20 Index&Poland\\
XU100&Instanbul Stock Exchange National 100 Index&Turkey\\
\bottomrule
\end{tabular}
\end{table}

\begin{table}[htbp]
\centering
\caption{Descriptive statistics for the analyzed indices}
\label{tab2}
\footnotesize
\begin{tabular}{||c|cccc|cc|cc||}
\toprule \toprule
Index&mean&min&max&SD&skewness&ex. kurtosis&KPSS&p-value\\
\midrule \midrule
AEX&-0.0003&-0.0959&0.1003&0.0157&-0.0183&6.1531&0.1084&$>0.05$\\
ASE&-0.0006&-0.1021&0.1343&0.0169&-0.0697&5.0812&0.3531&$>0.05$\\
ATX&0.0002&-0.1025&0.1202&0.0150&-0.3410&8.2241&0.3141&$>0.05$\\
BEL20&-0.0001&-0.0832&0.0933&0.0135&0.0694&6.7098&0.1381&$>0.05$\\
BSE&0.0004&-0.1181&0.1599&0.0170&-0.1630&6.2487&0.1900&$>0.05$\\
BUSP&0.0004&-0.1210&0.1368&0.0193&-0.0641&4.5410&0.1229&$>0.05$\\
BUX&0.0004&-0.1265&0.1318&0.0169&-0.1105&6.3117&0.2860&$>0.05$\\
CAC&-0.0002&-0.0947&0.1060&0.0154&0.0594&5.3189&0.0944&$>0.05$\\
CSE&0.0008&-0.1391&0.1770&0.0152&0.2208&25.7090&1.2088&$<0.01$\\
DAX&-0.0001&-0.0887&0.1080&0.0159&0.0025&4.7729&0.1681&$>0.05$\\
DJI&0.0000&-0.0820&0.1051&0.0126&-0.0089&7.8817&0.0647&$>0.05$\\
FTSE&-0.0001&-0.0927&0.0938&0.0129&-0.1309&6.4856&0.1222&$>0.05$\\
HEX&-0.0003&-0.1441&0.1344&0.0193&-0.1933&5.2159&0.1886&$>0.05$\\
HIS&0.0001&-0.1770&0.1341&0.0166&-0.2283&12.5630&0.1306&$>0.05$\\
IBC&0.0008&-0.2066&0.1453&0.0155&-0.4151&25.8530&0.2665&$>0.05$\\
IGBM&-0.0001&-0.1875&0.1840&0.0153&0.0833&20.5300&0.1272&$>0.05$\\
IGRA&0.0008&-0.1144&0.1282&0.0147&-0.3550&10.3010&0.3896&$>0.05$\\
IPC&0.0005&-0.0727&0.1044&0.0144&0.0515&4.3402&0.1295&$>0.05$\\
IPSA&0.0007&-0.0717&0.1180&0.0108&-0.0140&10.7400&0.1663&$>0.05$\\
JKSE&0.0006&-0.1095&0.0762&0.0150&-0.6570&6.1905&0.3397&$>0.05$\\
JSE&0.0006&-0.0758&0.0683&0.0135&-0.1786&3.2503&0.2009&$>0.05$\\
KFX&0.0002&-0.1172&0.0950&0.0137&-0.2594&5.7183&0.0939&$>0.05$\\
KLSE&0.0002&-0.1122&0.0537&0.0092&-1.1810&15.4970&0.1591&$>0.05$\\
KS11&0.0002&-0.1212&0.1128&0.0174&-0.4309&4.5849&0.1617&$>0.05$\\
MERVAL&0.0006&-0.1295&0.1612&0.0214&-0.1235&5.6617&0.1006&$>0.05$\\
MIBTEL&0.0002&-0.0771&0.0683&0.0108&-0.3979&5.7820&0.4301&$>0.05$\\
NASD&-0.0002&-0.1029&0.1116&0.0175&-0.1624&3.9587&0.2958&$>0.05$\\
NIKKEI&-0.0003&-0.1211&0.1324&0.0158&-0.3633&7.3242&0.1252&$>0.05$\\
NYA&0.0002&-0.1023&0.1153&0.0140&-0.4233&10.5210&0.1514&$>0.05$\\
PSE&-0.0001&-0.1860&0.2929&0.0162&1.8252&67.2470&0.2770&$>0.05$\\
PX50&0.0003&-0.1619&0.1236&0.0154&-0.6011&15.4230&0.4121&$>0.05$\\
SAX&0.0007&-0.0882&0.0711&0.0120&-0.0481&6.5294&0.5215&$>0.05$\\
SET&0.0000&-0.2211&0.1058&0.0158&-1.8111&26.2170&0.2975&$>0.05$\\
SPX&-0.0001&-0.0947&0.1096&0.0134&-0.1842&8.1808&0.0958&$>0.05$\\
SSEC&0.0002&-0.1200&0.0903&0.0168&-0.2784&4.7064&0.1461&$>0.05$\\
SSMI&-0.0001&-0.0811&0.1079&0.0127&0.0331&6.2488&0.0918&$>0.05$\\
STRAITS&0.0000&-0.2685&0.1406&0.0137&-2.2597&56.9590&0.1989&$>0.05$\\
TA100&0.0003&-0.0734&0.0978&0.0141&-0.1535&3.2977&0.1157&$>0.05$\\
TSE&0.0001&-0.0979&0.0937&0.0122&-0.6630&8.9915&0.0782&$>0.05$\\
WIG20&0.0004&-0.0886&0.3322&0.0185&2.6452&52.0680&0.1909&$>0.05$\\
XU100&0.0004&-0.1334&0.1749&0.0230&0.0039&4.5896&0.1105&$>0.05$\\
\bottomrule
\end{tabular}
\end{table}

\newpage

\section*{References}
\bibliography{EMH}

\begin{thebibliography}{}

\bibitem[\protect\citeauthoryear{Alessio, Carbone, Castelli, and
  Frappietro}{Alessio et~al.}{2002}]{Alessio2002}
Alessio, E., A.~Carbone, G.~Castelli, and V.~Frappietro (2002).
\newblock Second-order moving average and scaling of stochastic time series.
\newblock {\em European Physical Journal B\/}~{\em 27}, 197--200.

\bibitem[\protect\citeauthoryear{Alexeev and Tapon}{Alexeev and
  Tapon}{2011}]{Alexeev2011}
Alexeev, V. and F.~Tapon (2011).
\newblock Testing weak form efficiency on the {T}oronto {S}tock {E}xchange.
\newblock {\em Journal of {E}mpirical {F}inance\/}~{\em 18}, 661--691.

\bibitem[\protect\citeauthoryear{Alvarez-Ramirez, Cisneros, Ibarra-Valdez, and
  Soriano}{Alvarez-Ramirez et~al.}{2002}]{Alvarez-Ramirez2002}
Alvarez-Ramirez, J., M.~Cisneros, C.~Ibarra-Valdez, and A.~Soriano (2002).
\newblock Multifractal {H}urst analysis of crude oil prices.
\newblock {\em Physica A\/}~{\em 313}, 651--670.

\bibitem[\protect\citeauthoryear{Barabasi, Szepfalusy, and Vicsek}{Barabasi
  et~al.}{1991}]{Barabasi1991}
Barabasi, A.-L., P.~Szepfalusy, and T.~Vicsek (1991).
\newblock Multifractal spectre of multi-affine functions.
\newblock {\em Physica A\/}~{\em 178}, 17--28.

\bibitem[\protect\citeauthoryear{Barkoulas and Baum}{Barkoulas and
  Baum}{1996}]{Barkoulas1996}
Barkoulas, J. and C.~Baum (1996).
\newblock Long-term dependence in stock returns.
\newblock {\em Economics Letters\/}~{\em 53}, 253--259.

\bibitem[\protect\citeauthoryear{Barunik, Aste, Di~Matteo, and Liu}{Barunik
  et~al.}{2012}]{Barunik2012}
Barunik, J., T.~Aste, T.~Di~Matteo, and R.~Liu (2012).
\newblock Understanding the source of multifractality in financial markets.
\newblock {\em Physica A\/}~{\em 391}, 4234--4251.

\bibitem[\protect\citeauthoryear{Barunik and Kristoufek}{Barunik and
  Kristoufek}{2010}]{Barunik2010}
Barunik, J. and L.~Kristoufek (2010).
\newblock On {H}urst exponent estimation under heavy-tailed distributions.
\newblock {\em Physica A\/}~{\em 389(18)}, 3844--3855.

\bibitem[\protect\citeauthoryear{Cajueiro and Tabak}{Cajueiro and
  Tabak}{2004a}]{Cajueiro2004b}
Cajueiro, D. and B.~Tabak (2004a).
\newblock Evidence of long range dependence in {Asian} equity markets: the role
  of liquidity and market restrictions.
\newblock {\em Physica A\/}~{\em 342}, 656--664.

\bibitem[\protect\citeauthoryear{Cajueiro and Tabak}{Cajueiro and
  Tabak}{2004b}]{Cajueiro2004}
Cajueiro, D. and B.~Tabak (2004b).
\newblock The {H}urst exponent over time: testing the assertion that emerging
  markets are becoming more efficient.
\newblock {\em Physica A\/}~{\em 336}, 521--537.

\bibitem[\protect\citeauthoryear{Cajueiro and Tabak}{Cajueiro and
  Tabak}{2004c}]{Cajueiro2004a}
Cajueiro, D. and B.~Tabak (2004c).
\newblock Ranking efficiency for emerging markets.
\newblock {\em Chaos, Solitons and Fractals\/}~{\em 22}, 349--352.

\bibitem[\protect\citeauthoryear{Cajueiro and Tabak}{Cajueiro and
  Tabak}{2005}]{Cajueiro2005}
Cajueiro, D. and B.~Tabak (2005).
\newblock Ranking efficiency for emerging equity markets {II}.
\newblock {\em Chaos, Solitons and Fractals\/}~{\em 23}, 671--675.

\bibitem[\protect\citeauthoryear{Cajueiro and Tabak}{Cajueiro and
  Tabak}{2007}]{Cajueiro2007}
Cajueiro, D. and B.~Tabak (2007).
\newblock Long-range dependence and multifractality in the term structure of
  {LIBOR} interest rates.
\newblock {\em Physica A\/}~{\em 373}, 603--614.

\bibitem[\protect\citeauthoryear{Carbone, Castelli, and Stanley}{Carbone
  et~al.}{2004}]{Carbone2004}
Carbone, A., G.~Castelli, and H.~Stanley (2004).
\newblock Time-dependent {H}urst exponent in financial time series.
\newblock {\em Physica A\/}~{\em 344}, 267–271.

\bibitem[\protect\citeauthoryear{Chan, Hall, and Poskitt}{Chan
  et~al.}{1996}]{Chan1996}
Chan, G., P.~Hall, and D.~Poskitt (1996).
\newblock Periodogram-based estimators of fractal properties.
\newblock {\em Annals of Statistics\/}~{\em 23(5)}, 1684--1711.

\bibitem[\protect\citeauthoryear{Charles, Darne, and Fouilloux}{Charles
  et~al.}{2011}]{Charles2011}
Charles, A., O.~Darne, and J.~Fouilloux (2011).
\newblock {Testing the martingale difference hypothesis in $CO_2$ emission
  allowances}.
\newblock {\em Economic Modelling\/}~{\em 28}, 27--35.

\bibitem[\protect\citeauthoryear{Chong, Lam, and Yan}{Chong
  et~al.}{2011}]{Chong2011}
Chong, T., T.-H. Lam, and I.~Yan (2011).
\newblock Is the {C}hinese stock market really inefficient?
\newblock {\em China Economic Review\/}~{\em 23(1)}, 122–137.

\bibitem[\protect\citeauthoryear{Di~Matteo}{Di~Matteo}{2007}]{DiMatteo2007}
Di~Matteo, T. (2007).
\newblock Multi-scaling in finance.
\newblock {\em Quantitatice Finance\/}~{\em 7(1)}, 21--36.

\bibitem[\protect\citeauthoryear{Di~Matteo, Aste, and Dacorogna}{Di~Matteo
  et~al.}{2003}]{DiMatteo2003}
Di~Matteo, T., T.~Aste, and M.~Dacorogna (2003).
\newblock Scaling behaviors in differently developed markets.
\newblock {\em Physica A\/}~{\em 324}, 183--188.

\bibitem[\protect\citeauthoryear{Di~Matteo, Aste, and Dacorogna}{Di~Matteo
  et~al.}{2005}]{DiMatteo2005}
Di~Matteo, T., T.~Aste, and M.~Dacorogna (2005).
\newblock Long-term memories of developed and emerging markets: Using the
  scaling analysis to characterize their stage of development.
\newblock {\em Journal of Banking \& Finance\/}~{\em 29}, 827--851.

\bibitem[\protect\citeauthoryear{Eom, Oh, and Jung}{Eom et~al.}{2008}]{Eom2008}
Eom, C., G.~Oh, and W.-S. Jung (2008).
\newblock Relationship between efficiency and predictability in stock price
  change.
\newblock {\em Physica A\/}~{\em 387}, 5511--5517.

\bibitem[\protect\citeauthoryear{Fama}{Fama}{1970}]{Fama1970}
Fama, E. (1970).
\newblock {Efficient Capital Markets: A Review of Theory and Empirical Work}.
\newblock {\em Journal of Finance\/}~{\em 25}, 383--417.

\bibitem[\protect\citeauthoryear{Fama}{Fama}{1991}]{Fama1991}
Fama, E. (1991).
\newblock {Efficient Capital Markets: II}.
\newblock {\em Journal of Finance\/}~{\em 46(5)}, 1575--1617.

\bibitem[\protect\citeauthoryear{Genton}{Genton}{1998}]{Genton1998}
Genton, M. (1998).
\newblock Highly robust variogram estimation.
\newblock {\em Mathematical Geology\/}~{\em 30}, 213--221.

\bibitem[\protect\citeauthoryear{Gneiting and Schlather}{Gneiting and
  Schlather}{2004}]{Gneiting2004}
Gneiting, T. and M.~Schlather (2004).
\newblock {Stochastic Models That Separate Fractal Dimension and the Hurst
  Effect.}
\newblock {\em SIAM Review\/}~{\em 46(2)}, 269--282.

\bibitem[\protect\citeauthoryear{Gneiting, Sevcikova, and Percival}{Gneiting
  et~al.}{2010}]{Gneiting2010}
Gneiting, T., H.~Sevcikova, and D.~Percival (2010).
\newblock Estimators of fractal dimension: Assessing the roughness of time
  series and spatial data.
\newblock Technical report, Department of Statistics, University of Washington.

\bibitem[\protect\citeauthoryear{Hall and Wood}{Hall and Wood}{1993}]{Hall1993}
Hall, P. and A.~Wood (1993).
\newblock On the performane of box-counting estimators of fractal dimension.
\newblock {\em Biometrika\/}~{\em 80(1)}, 246--252.

\bibitem[\protect\citeauthoryear{Hurst}{Hurst}{1951}]{Hurst1951}
Hurst, H. (1951).
\newblock Long term storage capacity of reservoirs.
\newblock {\em Transactions of the American Society of Engineers\/}~{\em 116},
  770--799.

\bibitem[\protect\citeauthoryear{Kantelhardt, Zschiegner, Koscielny-Bunde,
  Bunde, Havlin, and Stanley}{Kantelhardt et~al.}{2002}]{Kantelhardt2002}
Kantelhardt, J., S.~Zschiegner, E.~Koscielny-Bunde, A.~Bunde, S.~Havlin, and
  E.~Stanley (2002).
\newblock {Multifractal Detrended Fluctuation Analysis of Nonstationary Time
  Series}.
\newblock {\em Physica A\/}~{\em 316(1-4)}, 87--114.

\bibitem[\protect\citeauthoryear{Kristoufek}{Kristoufek}{2010a}]{Kristoufek2010}
Kristoufek, L. (2010a).
\newblock On spurious anti-persistence in the {US} stock indices.
\newblock {\em Chaos, Solitons and Fractals\/}~{\em 43}, 68--78.

\bibitem[\protect\citeauthoryear{Kristoufek}{Kristoufek}{2010b}]{Kristoufek2010a}
Kristoufek, L. (2010b).
\newblock Rescaled range analysis and detrended fluctuation analysis: {F}inite
  sample properties and confidence intervals.
\newblock {\em {AUCO} {C}zech {E}conomic {R}eview\/}~{\em 4}, 236--250.

\bibitem[\protect\citeauthoryear{Kristoufek}{Kristoufek}{2012a}]{Kristoufek2012a}
Kristoufek, L. (2012a).
\newblock {Fractal Markets Hypothesis and the Global Financial Crisis: Scaling,
  Investment Horizons and Liquidity}.
\newblock {\em Advances in Complex Systems\/}~{\em 15(6)}, 1250065.

\bibitem[\protect\citeauthoryear{Kristoufek}{Kristoufek}{2012b}]{Kristoufek2012}
Kristoufek, L. (2012b).
\newblock How are rescaled range analyses affected by different memory and
  distributional properties? {A} {M}onte {C}arlo study.
\newblock {\em Physica A\/}~{\em 391}, 4252--4260.

\bibitem[\protect\citeauthoryear{Lam and Tam}{Lam and Tam}{2011}]{Lam2011}
Lam, K. and L.~Tam (2011).
\newblock Liquidity and asset pricing: Evidence from {Hong Kong} stock market.
\newblock {\em Journal of Banking \& Finance\/}~{\em 35}, 2217--2230.

\bibitem[\protect\citeauthoryear{Lim, Brooks, and Kim}{Lim
  et~al.}{2008}]{Lim2008}
Lim, K.-P., R.~Brooks, and J.~Kim (2008).
\newblock Financial crisis and stock market efficiency: {E}mpirical evidence
  from {A}sian countries.
\newblock {\em International Review of Financial Analysis\/}~{\em 17},
  571--591.

\bibitem[\protect\citeauthoryear{Malkiel}{Malkiel}{2003}]{Malkiel2003}
Malkiel, B. (2003).
\newblock The efficient market hypothesis and its critics.
\newblock {\em Journal of Economic Perspectives\/}~{\em 17(1)}, 59–82.

\bibitem[\protect\citeauthoryear{Mandelbrot}{Mandelbrot}{1982}]{Mandelbrot1982}
Mandelbrot, B. (1982).
\newblock {\em The fractal geometry of nature}.
\newblock W. H. Freeman Press.

\bibitem[\protect\citeauthoryear{Mandelbrot and van Ness}{Mandelbrot and van
  Ness}{1968}]{Mandelbrot1968}
Mandelbrot, B. and J.~van Ness (1968).
\newblock Fractional {B}rownian motions, fractional noises and applications.
\newblock {\em {SIAM} {R}eview\/}~{\em 10(4)}, 422--437.

\bibitem[\protect\citeauthoryear{Morales, Di~Matteo, Gramatica, and
  Aste}{Morales et~al.}{2012}]{Morales2012}
Morales, R., T.~Di~Matteo, R.~Gramatica, and T.~Aste (2012).
\newblock Dynamical generalized {H}urst exponent as a tool to monitor unstable
  periods in financial series.
\newblock {\em Physica A\/}~{\em 391}, 3180--3189.

\bibitem[\protect\citeauthoryear{Neves, Oliveira, Peres, Moreira, Moriel,
  de~Godoy, and Murta~Junior}{Neves et~al.}{2011}]{Neves2011}
Neves, L., F.~Oliveira, F.~Peres, R.~Moreira, A.~Moriel, M.~de~Godoy, and
  L.~Murta~Junior (2011).
\newblock Maximum entropy, fractal dimension and lacunarity in quantification
  of cellular rejection in myocardial biopsy of patients submitted to heart
  transplantation.
\newblock {\em Journal of Physics: Conference Series\/}~{\em 285}, 012032.

\bibitem[\protect\citeauthoryear{Panas}{Panas}{2001}]{Panas2001}
Panas, E. (2001).
\newblock Estimating fractal dimension using stable distributions and exploring
  long memory through {ARFIMA} models in {Athens Stock Exchange}.
\newblock {\em Applied Financial Economics\/}~{\em 11}, 395--402.

\bibitem[\protect\citeauthoryear{Peng, Buldyrev, Goldberger, Havlin, Simons,
  and Stanley}{Peng et~al.}{1993}]{Peng1993}
Peng, C., S.~Buldyrev, A.~Goldberger, S.~Havlin, M.~Simons, and H.~Stanley
  (1993).
\newblock Finite-size effects on long-range correlations: Implications for
  analyzing {DNA} sequences.
\newblock {\em Physical Review E\/}~{\em 47(5)}, 3730--3733.

\bibitem[\protect\citeauthoryear{Peng, Buldyrev, Havlin, Simons, Stanley, and
  Goldberger}{Peng et~al.}{1994}]{Peng1994}
Peng, C., S.~Buldyrev, S.~Havlin, M.~Simons, H.~Stanley, and A.~Goldberger
  (1994).
\newblock Mosaic organization of {DNA} nucleotides.
\newblock {\em Physical Review E\/}~{\em 49\/}(2), 1685--1689.

\bibitem[\protect\citeauthoryear{Percival and Walden}{Percival and
  Walden}{2000}]{Percival2000}
Percival, D. and A.~Walden (2000).
\newblock {\em Wavelet Methods for Time Series Analysis}.
\newblock Cambridge University Press.

\bibitem[\protect\citeauthoryear{Podobnik, Fu, Jagric, Grosse, and
  Stanley}{Podobnik et~al.}{2006}]{Podobnik2006}
Podobnik, B., D.~Fu, T.~Jagric, I.~Grosse, and H.~E. Stanley (2006).
\newblock Fractionally integrated process for transition economics.
\newblock {\em Physica A\/}~{\em 362}, 465--470.

\bibitem[\protect\citeauthoryear{Vandewalle, Ausloos, and Boveroux}{Vandewalle
  et~al.}{1997}]{Vandewalle1997}
Vandewalle, N., M.~Ausloos, and P.~Boveroux (1997).
\newblock {Detrended Fluctuation Analysis of the Foreign Exchange Market}.
\newblock In {\em Econophysic Workshop}, Budapest, Hungary.

\bibitem[\protect\citeauthoryear{Weron}{Weron}{2002}]{Weron2002}
Weron, R. (2002).
\newblock Estimating long-range dependence: finite sample properties and
  confidence intervals.
\newblock {\em Physica A\/}~{\em 312}, 285–--299.

\bibitem[\protect\citeauthoryear{Wright}{Wright}{2001}]{Wright2001}
Wright, J.~H. (2001).
\newblock Long memory in emerging market stock returns.
\newblock {\em Emerging Markets Quarterly\/}~{\em 5}, 50--55.

\end{thebibliography}
\bibliographystyle{chicago}

\end{document}